\documentstyle[11pt,amsfonts]{article}

\newcommand{\be}{\begin{equation}}
\newcommand{\ee}{\end{equation}}
\newcommand{\bea}{\begin{array}}
\newcommand{\ea}{\end{array}}
\newcommand{\beqa}{\begin{eqnarray}}
\newcommand{\eeqa}{\end{eqnarray}}
\newcommand{\bean}{\begin{eqnarray*}}
\newcommand{\eean}{\end{eqnarray*}}

\def\up#1{\leavevmode \raise.16ex\hbox{#1}}

\newcommand{\gapproxeq}{\lower
 .7ex\hbox{$\;\stackrel{\textstyle >}{\sim}\;$}}
\newcommand{\lapproxeq}{\lower .7ex\hbox{$\;\stackrel
{\textstyle <}{\sim}\;$}}

\newcounter{appendice}

\def\thebibliography#1{{\bf REFERENCES\markboth
 {REFERENCES}{REFERENCES}}\list
 {[\arabic{enumi}]}{\settowidth\labelwidth{[#1]}\leftmargin\labelwidth
 \advance\leftmargin\labelsep
 \usecounter{enumi}}
 \def\newblock{\hskip .11em plus .33em minus -.07em}
 \sloppy
 \sfcode`\.=1000\relax}

\begin{document}
\vskip 1cm
\centerline{ \LARGE Particle Dynamics  on Snyder space }

\vskip 2cm

\centerline{{\sc Lei Lu}\footnote{llv1@crimson.ua.edu }  and    {\sc A. Stern }\footnote{astern@bama.ua.edu}   }

\vskip 5mm

\centerline{  Department of Physics and Astronomy, University of Alabama,
Tuscaloosa, Al 35487, United States}
\vskip 2cm

\vspace*{5mm}

\normalsize
\centerline{\bf ABSTRACT}
We examine  Hamiltonian formalism on Euclidean Snyder space.  The latter corresponds to a lattice in the quantum theory.
For any given dynamical system, it may not be possible to identify  time with a real number  parametrizing the evolution in the quantum theory.  
The alternative requires the introduction of a dynamical time operator.  We obtain the dynamical time operator for the relativistic (nonrelativistic) particle, and use it to construct the generators of Poincar\'e (Galilei) group
on Snyder space.

\vskip 4 cm 
\newpage

\section{Introduction}

Snyder showed long ago that a spatial lattice can be  consistent with the continuous symmetries of space-time through the construction of a covariant noncommutative algebra.\cite{Snyder:1946qz}  The algebra  is generated by two Lorentz vectors:  the time-space  coordinate $\hat x_\mu$, and the energy-momentum   $\hat p_\mu$, $\mu=0,1,2,3$.  The spatial lattice of Snyder, or `Snyder space', results from the fact that  the spatial coordinates $\hat x_i$, $i=1,2,3$, have discrete spectra.  It differs from  a classical lattice because  only one coordinate  can be determined in a measurement due to the noncommutativity.  In a previous article,\cite{us}  we found that  two distinct representations of the Snyder algebra are possible. They are characterized by an $SU(2)$ quantum number, which takes on integer values for one representation and half-integer values for the other.  Continuous symmetry transformations are unitarily implementable on the lattice. The continuous transformations  examined in \cite{us} were generated by the momenta and angular momenta, and are associated with the  translation\footnote{As was shown in \cite{us}, they are not translations from one point on the lattice to another, but rather  are translations in a particular  continuous basis of the Hilbert space. } and rotation group, respectively. In this article  we shall  demonstrate, among other things, how to write down the Hermitean generators for  the full Poincar\'e  group (Galilei group).  This is facilitated  by examining the Hamiltonian dynamics of a relativistic  particle  (nonrelativistic  particle) on Snyder space.

  Covariance is problematic in the standard Hamiltonian description of a relativistic particle since one treats time  differently from the spatial coordinates. It is  then not surprising that the Lorentz covariant algebra of Snyder is inconsistent with the Hamiltonian description of a relativistic particle.   Various tactics can be taken to get around this issue, and they have been employed  in previous articles.  These include introducing additional degrees of freedom\cite{Jaroszkiewicz},  such as an additional time variable\cite{Romero:2004er}, or modifying the particle action\cite{Banerjee:2006wf}. 
 The approach adopted in this article does not require either introducing additional degrees of freedom or  deforming the dynamics.  Instead, as in the standard Hamiltonian description,  we can drop the insistence  on covariance, and still have a  consistent relativistic particle dynamics.  The latter is written down on the Euclidean subalgebra of the Snyder algebra, generated by  the  spatial components of the four vectors, i.e., $\hat x_i$ and $\hat p_i$.\footnote{Some aspects of nonrelativistic dynamics on this space were examined previously in \cite{Mignemi:2011gr}.}  Although, $\hat p^0$ and $\hat x^0$,  no longer appear as independent generators of the algebra in this approach, the notion of energy and time re-appear with the introduction of particle dynamics.  One can then derive algebraic properties for these quantities.  This is straightforward  for the operator $\hat p^0$ associated with the energy, once we define it  by the mass shell condition. 

 Different approaches can be taken with regard to   the time. 
  If one insists, as usual, that time is a real parameter associated with the evolution of the system,  then  $\hat p^0$ cannot be  the particle Hamiltonian on Snyder space because it will not in general generate time evolution.  Rather, the Hamiltonian is some function of $\hat p^0$.  This then implies nonstandard energy-momentum dispersion relations,  a feature in common with   double special relativity\cite{AmelinoCamelia:2010pd}.  
On the other hand, as  we argue here, nonstandard energy-momentum dispersion relations need not be a  signature of Snyder space. 
One can retain the conventional energy-momentum dispersion relation upon re-introducing a  time  operator. The time operator  in this case is not an independent generator of the algebra, but rather a function of the operators $\hat x_i$ and $\hat p_i$, along  with the evolution parameter denoted by $\lambda$.  For this reason we refer to it as a dynamical time operator.  The operator is only defined up to constants of the motion, which are not in general central in the algebra.  For several examples we  determine the definition of the time  by requiring it to be a gauge fixing condition in a reparametrization invariant formulation of the dynamics.  The choice for the gauge fixing is such that the Euclidean Snyder algebra is realized by the Dirac brackets for the constrained Hamiltonian system.  The examples considered in this article are the relativistic and nonrelativistic free particle, along with one-dimensional conservative dynamics.  The relativistic (nonrelativistic) free particle action that we start from is Lorentz (Galilei) invariant, as well as reparametrization invariant.  We are then able to construct  Lorentz (Galilei) boost operators from the Euclidean Snyder algebra which satisfy the Poincar\'e (Galilei) algebra.

The outline of this article is as follows.
In Section 2 we show  that the covariant Snyder algebra  is inconsistent with the mass shell condition. We then argue that dynamics can be consistently formulated on the Euclidean subalgebra. In Section 3, we write down the symplectic two-form associated with the Euclidean Snyder algebra and develop Hamiltonian  dynamics on the resulting phase space. Time is treated as a commuting number in this section, while dynamical time operators are constructed in Section 4.  We do this for the three examples of the nonrelativistic and  free relativistic particle, and the one-dimensional conservative system.
Momentum-dependent eigenfunctions of the time operator, along with space-time symmetry generators,   are obtained for the free particle examples.  Concluding remarks are made in Section 5.

\section{Covariance lost}

 The Lorentz covariant algebra of Snyder  is defined by the commutation relations\cite{Snyder:1946qz}
\beqa [\hat x_\mu,\hat x_\nu] &=&\frac i{\Lambda^2}(\hat x_\mu \hat p_\nu-\hat x_\nu \hat p_\mu)\cr& & \cr [\hat x_\mu,\hat p_\nu] &=&i\biggl(\eta_{\mu\nu} +\frac{\hat p_\mu\hat p_\nu}{\Lambda^2}\biggr)\cr & &\cr [\hat p_\mu,\hat p_\nu] &=&0\label{lrntzcvrnt}\eeqa 
where $\mu,\nu,...=0,1,2,3$ and we choose $[\eta_{\mu\nu}]$= diag$(-1,1,1,1)$ for the Minkowski metric. $\Lambda\ne 0$ is the deformation parameter, which has  units of energy.  The  canonical  commutation relations of  Minkowski space-time coordinates with the four-momenta are recovered in the  limit  $\Lambda\rightarrow\infty$.   The algebra  (\ref{lrntzcvrnt}) was originally obtained starting from the de Sitter group acting on de Sitter momentum space, and then projecting to Minkowski space-time.  The projection involves a particular coordinatization of de Sitter space by the four-momentum $\hat p_\mu$, and an identification of  four of the de Sitter group generators with the space-time coordinates $\hat x_\mu$.  Following that approach one arrives at an upper bound on the mass, $-\hat p^\mu\hat p_\mu\le\Lambda^2$, but this bound is not required if we instead simply postulate the commutation relations (\ref{lrntzcvrnt}) from the start.

 The  canonical  commutation relations of  Minkowski space-time coordinates and the four-momenta (corresponding to the limit $\Lambda\rightarrow\infty$)  are  inconsistent with the mass shell condition $\hat p^\mu\hat p_\mu +m^2=0$ defining a relativistic free particle.  This is since $\hat p^\mu\hat p_\mu $ is not central in the algebra. It is then not surprising that the  algebraic relations (\ref{lrntzcvrnt}) are  also inconsistent with the mass shell condition. This is   since
\be  [\hat x_\mu,\hat p^\nu\hat p_\nu]=2i\hat p_\mu\Bigl( 1+\frac{\hat p^\rho\hat p_\rho}{\Lambda^2}\Bigr)\ee
Just as one should not  insist on Lorentz covariance when writing down the Hamiltonian formalism for a relativistic particle  using the  canonical  commutation relations, one should not  insist on Lorentz covariance when writing down relativistic particle dynamics on Snyder space. 

Instead, Hamiltonian dynamics for a particle moving in three dimensions is consistently written down on a  six-dimensional phase space.  By Darboux's theorem we can take the latter to be  the Euclidean subalgebra of the Snyder algebra, at least classically.  At the quantum level, the algebra is generated by the three spatial components of the coordinates and momenta, $\hat x_i$ and  $\hat p_i$, $i=1,2,3$, and defined by the  commutation relations 
\beqa [\hat x_i,\hat x_j] &=&\frac i{\Lambda^2}\epsilon_{ijk}\hat L_k\cr& & \cr [\hat x_i,\hat p_j] &=&i\biggl(\delta_{ij} +\frac{\hat p_i\hat p_j}{\Lambda^2}\biggr)\cr & &\cr [\hat p_i,\hat p_j] &=&0\label{snydrsubalg}\eeqa 
where  $\hat L_i=\epsilon_{ijk}\hat x_j \hat p_k$ are the angular momenta.
 The three-momenta   $\hat p_i$ are simultaneously diagonalizable and have continuous eigenvalues $p_i$.
Snyder gave a  differential operator representation for $\hat x_i$ acting on the space of complex functions  $\{\phi,\psi,...\}$ of the momentum $\vec p=(p_1,p_2,p_3)$.  It is:\footnote{This was generalized to a one parameter family of  representations in \cite{us}.}
\beqa \hat x_i&\rightarrow &i \frac{\partial}{\partial p_i} +\frac{ ip_i p_j }{\Lambda^2} \frac{\partial}{\partial p_j }
\label{noncandifrep}\;\eeqa
These operators are symmetric for the scalar product 
\be (\phi,\psi)=\int  d\mu(\vec p)\;\phi(\vec p)^*\psi(\vec p)\;,\label{sclrprdctpsc}\ee
using the measure
\be d\mu(\vec p) =\frac{ d^3p}{(1+ \frac{ \vec p^2}{\Lambda^2})^{2}} \label{measure}\;\ee 

Whereas  Snyder's Lorentz covariant algebra (\ref{lrntzcvrnt})
 possesses the independent  operators $\hat x^0$ and $\hat p^0$, associated respectively with the time and energy, from (\ref{snydrsubalg}) we can define analogous operators as functions of $\hat x^i$ and $\hat p^i$  using particle dynamics.  For the relativistic free  particle, the energy  is naturally defined by the mass shell condition
\be \hat p^0=\sqrt{ \hat p_i\hat p_i+m^2}\;\label{pzero}\ee
 Then 
\be [\hat x_i,\hat p^0]=i\,\frac{\hat p_i}{\hat p^0}\Bigl(1+\frac{\hat p_k\hat p_k}{\Lambda^2}\Bigr)\qquad\qquad  [\hat p_i,\hat p^0]=0\;,\label{crofp0wxipi}\ee 
which differs from the commutation relations for $\hat p^0$ postulated  in (\ref{lrntzcvrnt}).  In the limit $\Lambda\rightarrow \infty$, (\ref{crofp0wxipi}) implies the usual Heisenberg equation of motion for a relativistic particle upon identifying $\hat p^0$ with the Hamiltonian.

As mentioned in the introduction, the notion of time for a relativistic  particle can   be addressed in different ways.   It can be regarded as the evolution parameter associated with the relevant Hamiltonian $H_t$, or it can be an operator defined in terms of some evolution generated by the zeroth-component of the four-momentum $\hat p^0$.   Both of these approaches coincide, when one uses the Heisenberg commutation relations, i.e., $\Lambda\rightarrow \infty$, since then $H_t=\hat p^0$.  However, this is not true in general, and specifically for commutation relations (\ref{snydrsubalg}).  In the former approach, we avoid dealing with a noncommuting time operator.  Then $H_t$ is some deformation of  $\hat p^0$, which we examine in the following section.
The latter approach  re-introduces  a time operator, which here is a function of the fundamental operators $\hat x_i$ and $\hat p_i$, as well as the evolution parameter for the system, which we denote by $\lambda$.  As stated in the introduction, the function may be interpreted as a gauge fixing condition starting from a reparametrization invariant action.  The commutation relations for this dynamically defined time operator can then be derived using (\ref{snydrsubalg}).  The advantage of  this approach is that one can retain $\hat p^0$ as the evolution operator for the  relativistic particle [or $\vec p^2/(2m)$ as the Hamiltonian for the  nonrelativistic particle] and therefore preserve the standard energy-momentum dispersion relations.   This appraoch is examined in section four.

\section{Hamiltonian formalism on Snyder space}
\setcounter{equation}{0}

\subsection{General formalism}

Before discussing quantum dynamics on the Snyder algebra, we  examine classical Hamiltonian dynamics on the associated classical phase space.  The  phase space  is spanned by  $ x_i$ and  $ p_i$, corresponding to the classical analogues of 
 $\hat  x_i$ and  $\hat p_i$.
The   symplectic two-form for the system is given by
\be \omega= dx_i\wedge dp_i -\frac 12 d(x_i p_i)\wedge d\ln(\vec p^2+\Lambda^2) \;\label{smplct2frm}\ee
To see this, we can invert the symplectic matrix to obtain the associated Poisson brackets.  The inverse exists because the determinant is   $(1+\vec p^2/\Lambda^2)^{-2}$, which  is nonvanishing for finite momentum.  The resulting Poisson brackets  are
\beqa \{ x_i, x_j\} &=&\frac 1{\Lambda^2}\;\epsilon_{ijk} L_k\;,\qquad\quad L_i=\epsilon_{ijk}x_jp_k\cr & &\cr  \{ x_i, p_j\} &=&\delta_{ij} +\frac{ p_ip_j}{\Lambda^2}\cr & &\cr \{ p_i, p_j\} &=&0\;,\label{classsnydr}\eeqa which
 are the classical analogues of the commutation relations (\ref{snydrsubalg}).  The second term in (\ref{smplct2frm}) drops out in the limit $\Lambda\rightarrow \infty$, and we recover the canonical  symplectic two-form  \be \omega_0=dx_i\wedge dp_i\;\label{cnclsmp2frm}\ee More generally, the Darboux map from   (\ref{smplct2frm}) to the canonical symplectic two-form (here written as $ dq_i\wedge dp_i$) 
is given by 
\be x_i\rightarrow q_i=x_i -\frac  { x_jp_jp_i}{\Lambda^2+\vec p^2} \;,\label{darboux}\ee  while the momenta $ p_i$ are invariant under the map. 
Thus the variables $ q_i$ are canonically conjugate to  $ p_i$.  

For dynamics on the phase space spanned by $x_i$ and $p_i$,  we introduce   a Hamiltonian $H$.  It  generates evolution in some parameter $\lambda$.  In general,
 $\lambda$ need not be  the time $t$, but rather some monotonically increasing function of the time. The  Hamilton equations of motion 
\be i_{\Delta_\lambda} \omega =dH\;\label{crdfreeeom}\ee
  relate $H$ to the dynamical vector field
\be \Delta_\lambda =\frac{d x_i}{d\lambda} \frac\partial{\partial x_i}+
\frac{d p_i}{d\lambda} \frac\partial{\partial p_i}\ee  From the symplectic two-form (\ref{smplct2frm}),  one gets the following equations for the first derivatives of $H$, 
\beqa \frac{\partial H}{\partial x_i}&=&-\frac{d p_i}{d\lambda}+\frac{\frac d{d\lambda}(\vec p^2)\;p_i}{2(\vec p^2 +\Lambda^2)}\cr & &\cr  \frac{\partial H}{\partial p_i}&=&\frac{d x_i}{d\lambda}+\frac{\frac d{d\lambda}(\vec p^2)\,x_i   -2\frac d{d\lambda}(\vec x\cdot\vec p)\,p_i}{2(\vec p^2 +\Lambda^2)}\label{frstdrvH}\eeqa 
 Consistency relations can be obtained by computing second derivatives of $H$.

Aspects of Hamiltonian dynamics on the above phase space were examined previously in \cite{Mignemi:2011gr},\cite{Battisti:2008xy}, along with the  equations of motion following from sample Hamiltonians,  in particular,  the harmonic oscillator Hamiltonian \be H_{\rm ho}=\frac{{ p}^2}{2m}+\frac 12m\omega^2  x^2\, \label {sho}\ee
 The evolution  parameter $\lambda$ was equal to the time in these works. With such an identification, the dynamics generated by $H$ on  the  phase space associated with  symplectic two-form (\ref{smplct2frm}) is, in general, inequivalent to the dynamics generated by the same Hamiltonian on  canonical phase space, the latter defined by the symplectic two-form (\ref{cnclsmp2frm}). 
Then,  for example,  $H_{\rm ho}$ along with (\ref{smplct2frm}) does not describe a simple harmonic oscillator, but rather a deformed  oscillator.  Such a system was examined in \cite{Mignemi:2011gr}.  In this  article  our aim is not to deform, but rather, to preserve the dynamics of any given system while passing from the canonical symplectic two-form (\ref{cnclsmp2frm}) to (\ref{smplct2frm}).

To identify the  equations of motion (\ref{crdfreeeom}) with some given system, we must specify both  $H$ and $\lambda$.  Say that the   dynamics is Hamiltonian with respect to  the canonical symplectic two-form (\ref{cnclsmp2frm}).  That is, it is defined by the symplectic two-form  (\ref {cnclsmp2frm}) and some Hamiltonian
$H_0$, with the latter   generating  evolution in the time $t$.\footnote{More generally, it may be possible to have is a solution to (\ref{frstdrvH}) even when  no Hamiltonian description with respect to the canonical symplectic two-form $\omega_0$ is possible.} Thus the dynamics is given by the standard Hamilton equations
\be i_{\Delta_t} \omega_0 =dH_0\;,\ee
where 
 $ \Delta_t =\frac{d\lambda}{dt}\Delta_\lambda=\frac{d x_i}{dt} \frac\partial{\partial x_i}+
\frac{d p_i}{dt} \frac\partial{\partial p_i}$, or in terms of components, 
\be \frac{\partial H_0}{\partial x_i}=-\frac{d p_i}{dt}\qquad\quad \frac{\partial H_0}{\partial p_i}=\frac{d x_i}{dt} \label{cnnclHmeq} \ee
Substituting (\ref{cnnclHmeq}) into (\ref{frstdrvH}) gives
\beqa \frac{d\lambda}{dt}\frac{\partial H}{\partial x_i}&=&\frac{\partial H_0}{\partial x_i}-\frac{p_ip_j}{\vec p^2 +\Lambda^2}\;\frac{\partial H_0}{\partial x_j}\cr & &\cr \frac{d\lambda}{dt}\frac{\partial H}{\partial p_i}&=&\frac{\partial H_0}{\partial p_i}-\frac 1{\vec p^2 +\Lambda^2}\biggl((x_i p_j-x_jp_i) \frac{\partial H_0}{\partial x_j}  +p_ip_j\frac{\partial H_0}{\partial p_j} \biggr)\label{frstdrvHzro}\eeqa The conditions state that $H$ generates the same dynamics on the phase space manifold with symplectic two-form (\ref{smplct2frm}), as  $H_0$ generates on the  phase space manifold with the canonical symplectic two-form (\ref{cnclsmp2frm}). 
The equations of motion (\ref{frstdrvHzro}) are valid in any number of dimensions.  For a given $H_0$, one can  search for consistent solutions to $H$ and  $\frac{d\lambda}{dt}$ as functions of the phase space variables $x_i$ and $p_i$, and also $\lambda$.  
These solutions, if they exist, may not be  unique.  Below we shall obtain solutions in two simple cases:  the case of one-dimensional motion and the case of free particle motion.

 \subsection{One-dimensional systems}
 
  We first consider the case of motion in one dimension, by which we mean a    two-dimensional phase space  spanned by $x$ and $p$.  Then Eqs. (\ref{frstdrvHzro})  simplify to
\beqa \frac{d\lambda}{dt}\frac{\partial H}{\partial x}&=&\frac{\Lambda^2}{ p^2 +\Lambda^2}\;\frac{\partial H_0}{\partial x}\cr & &\cr \frac{d\lambda}{dt}\frac{\partial H}{\partial p}&=&\frac {\Lambda^2}{ p^2 +\Lambda^2} \frac{\partial H_0}{\partial p} \label{1dfrstdrvHzro}\eeqa
They are satisfied upon writing
\be H=f(H_0)\quad\qquad \frac{dt}{d\lambda}=\Bigl(1+\frac {p^2}{\Lambda^2}\Bigr)\,f'(H_0)\;,\label{gnrlslnfrHdrv}\ee
where $f(H_0)$ is a function of $H_0$, and $f'(H_0)$ is its derivative. 
 It remains to specify the evolution parameter $\lambda$.  Here we should require that $\frac{dt}{d\lambda}$ is everywhere positive for all physical particle trajectories.  Of course, this is not a concern  if we simply identify   $\lambda$ with the time $t$.  In that case, from  (\ref{gnrlslnfrHdrv}), we would need $f'(H_0)=(1+ {p^2}/{\Lambda^2})^{-1}$.
This can be satisfied when $H_0$ depends only on $p$, but it is not in general valid for an arbitrary $x$-dependent Hamiltonian, such as the case with a harmonic oscillator. 

Alternatively, instead of trying to identify $t$ with $\lambda$,  one can  set $H=H_0$, and then solve for $\lambda$ as a function of $x$, $ p$ and $t$.  This is always possible on the two-dimensional phase space.  In that case, one has
\be H=H_0 \qquad\quad \frac{dt}{d\lambda}=1+\frac { p^2}{\Lambda^2}\label{heqlhzro}\;,\ee 
	and it follows that $\frac{dt}{d\lambda}$ is everywhere positive.\footnote{We assume that $\Lambda^2>0$.} 
To obtain the relation between $t$ and $\lambda$ we should integrate (\ref{heqlhzro}) on the space of solutions: $x=x^{\xi_1,\xi_2,...}_{\rm sol}(t)$, $p=p^{\xi_1,\xi_2,...}_{\rm sol}(t)$, where ${\xi_1,\xi_2,...}$ label the solutions and correspond to constants of motion for the system.
Then  \be\lambda=\Lambda^2 \int^t\frac {dt'} {\Lambda^2+[p^{\xi_1,\xi_2,...}_{\rm sol}(t')]^2} \;,\label{ntgrlfrt}\ee from which one can in principle solve for $t$ as a function of $\lambda$ and the phase space variables. Thus $t$ is  dynamically determined, and  in general will have nontrivial  Poisson brackets.  However, the latter are not uniquely determined, since one can add a function of the constants of the motion ${\xi_1,\xi_2,...}$ to (\ref{ntgrlfrt}).  The constants of motion are not, in general,  central in the algebra, so the commutation relations for the time operator depends on this function. 
In section 4, we fix this arbitrariness by demanding that the symplectic structure (\ref{smplct2frm}) be derived starting from a particle action.  We require  the action to be reparametrization invariant.  The Snyder algebra is then realized by the resulting Dirac brackets for the system upon fixing a gauge condition. The gauge fixing condition then defines the time in terms of the evolution parameter.  (This procedure was previously carried out for the case of the relativistic free particle in \cite{Stern:2010ri}.  Here we also apply it to the nonrelativistic particle and one-dimensional conservative systems. )

\subsection{Free particles}

The problem of finding solutions to (\ref{frstdrvHzro}) in more than one dimension is more complicated - with the exception of the free particle case, which we now discuss. Here we assume that
 $H_0$ is independent of $x_i$.   (\ref{frstdrvHzro}) then implies that $H$ also only depends  on the momenta (and possibly $\lambda$) and  the dependence is such that 
\be   \frac{d\lambda}{dt}\;\frac{\partial H}{\partial p_i}= \biggl(\delta_{ij}-\frac {p_ip_j}{\ p^2 +\Lambda^2}\biggr)\frac{\partial H_0}{\partial p_j}\label{gnrlfrprtcl}\ee
Upon insisting on rotational  invariance, i.e., $H_0$ is a function  only of $p^2$, and  the equations simplify further to
\be \frac{d\lambda}{dt}\frac{\partial H}{\partial p_i}=\frac {\Lambda^2}{ p^2 +\Lambda^2} \frac{\partial H_0}{\partial p_i}\;
\ee
The solutions are once again (\ref{gnrlslnfrHdrv}). It again remains to fix $H$ and $\lambda$. Here given $H_0$ one can consistently set $t=\lambda$ and solve for $H$ as a function of $p^2$.  We do this explicitly below for the nonrelativistic and relativistic free particle. 

Starting with the  nonrelativistic free particle, with $ H_0=\frac{\vec p^2}{2m}$,  we get
\be  t=\lambda \qquad\quad
 H=H_t=\frac {\Lambda^2}{2m}\ln { \Bigl(1+ \frac{ \vec { p}^2}{\Lambda^2}\Bigr) }\,\label{nonrelham}\ee   up to constants of integration.
  Regarding $H_t$ as the energy, (\ref{nonrelham}) implies a non-standard  energy-momentum dispersion relation. Upon expanding $H_t$ in $1/\Lambda$, the standard nonrelativistic kinetic energy appears at zeroth order and the leading correction  is fourth order in momentum, \be H_t= \frac{\vec { p}^2}{2m} -\frac {(\vec {  p}^2)^2}{4m\Lambda^2}+\cdots\;\ee 

For the case of the relativistic free particle, we can choose  $H_0= p^0$, where
\be p^0
=\sqrt{  {\vec p}^2+m^2}\label{pzerocv}\ee  It leads to the equations of motion 
 $ {d\vec x}/{dt}=\vec p/p^0$. Here $t$ should be interpreted as the zero component of the space-time four vector. Up to constants of integration, the Hamiltonian $H$ which generates the motion on the  phase space associated with symplectic two-form (\ref{smplct2frm}) is
\be  t=\lambda\qquad\quad
 H =H_t=\left\{\matrix{\frac {\Lambda^2}{\sqrt{\Lambda^2 -m^2}}\;\tan^{-1}
\frac { p^0} {\sqrt{\Lambda^2 -m^2}}\;,\quad &m<\Lambda\cr &\cr \frac {\Lambda^2}{\sqrt{m^2-\Lambda^2 }}\;\tanh^{-1}
\frac { p^0} {\sqrt{m^2-\Lambda^2 }}\;,&\quad m>\Lambda\cr }\right.\;\;\label{relham}\ee  
(This result  appeared previously in \cite{Stern:2010ri}.) 
Regarding $H_t$ as the energy, we once again  get a nonstandard dispersion relation for the energy-momentum for finite $\Lambda$.
 One recovers  $ p^0$ from (\ref{relham}) in the  limit $\Lambda\rightarrow \infty$.  The leading $1/\Lambda^2$ correction  is \be \frac{\sqrt{ \vec { p}^2+m^2}}{3\Lambda^2}(2 m^2 -\vec { p}^2) \;,\ee which includes a $\frac 23 \frac {m^3}{\Lambda^2}$ correction to the rest mass energy. 

For both of the above free particle solutions, time remains a commutative parameter in the quantum theory. It, along with operator analogue $\hat H_t$ of the Hamiltonian $H_t$, defines the quantum evolution operator \be U(t,0)=e^{-i\hat H_tt}\label{evlnintm}\ee
From the evolution operator one recovers the usual equations for the  expectation value of the operators. For the nonrelativistic particle, one gets
\be \frac d{dt} <\hat x_i>= -i<[\hat x_i,\hat H_t]> =
\frac{<\hat p_i>}m  \qquad\quad  \frac d{dt} <\hat p_i>= 0\;,  \label{stndrdxpcxnr}\ee
while  for the relativistic particle 
\be \frac d{dt} <\hat x_i>= -i<[\hat x_i,\hat H_t]> =
<\frac{\hat p_i}{\hat p^0} >   \qquad\quad  \frac d{dt} <\hat p_i>= 0 \label{dmnxdt}   \;,  \ee  where
 $ \hat p^0$ is defined in (\ref{pzero}).

Unlike in the above, generic solutions to (\ref{frstdrvHzro}) will have $\frac {d\lambda}{dt}\ne 1$, and so then time cannot be identified with the evolution parameter, either in the classical or the quantum theory.  In that case, time is a function
of the position and momenta, as well as $\lambda$, in the classical theory, while it is 
 an operator in the corresponding quantum theory.  We examine this possibility, specifically for the case $H=H_0$, in the next section.  

\section{Dynamical time operators}
\setcounter{equation}{0}

In this section we choose $H$ to be the Hamiltonian $H_0$ which generates  evolution on the canonical phase space.  By so doing, we avoid  nonstandard energy-momentum dispersion relations, such as in (\ref{nonrelham}) and (\ref{relham}).   We obtain the time $t$  as a function
of $\vec x$, $\vec p$ and $\lambda$ in the classical theory for three examples, the free nonrelativistic particle,  the free relativistic particle and one-dimensional conservative systems.  We do this
by demanding that it corresponds to a gauge fixing condition for a reparametrization invariant  action.

\subsection{Nonrelativistic free particle}

Since the momentum $\vec p$ is  constant for free particle trajectories,   (\ref{heqlhzro}) implies that $t$  is a rescaling of  $\lambda$ by a constant factor.
From   (\ref{ntgrlfrt}) one gets
	\be \lambda=\frac{\Lambda^2 t+C}{\vec{  p}^2+\Lambda^2}\;,\label{lmdasntgrl}\ee along any trajectory.  
 $C$ is some function of the constants of motion, which here include the Galilei boosts
 \be\vec G=t\vec p-m\vec x\;,\label{vecG}\ee as well as $\vec p$ and $\vec L$.  Below we  shall show that if we make the choice $C=\vec p \cdot\vec G$, and hence\be t=\lambda +\frac {m\vec x\cdot \vec p}{\vec{ p}^2+\Lambda^2}\;,\label{ncnreltime} \ee
that we recover the full Galilei symmetry for the nonrelativistic free particle and, in particular, $ G_i$ satisfy the algebra of Galilei boosts.  

 Galilei invariance is, of course, insured if one starts with the usual free particle action $S^0_{\rm nr}=\int dt\; \frac m2 \Bigl(\frac{d\vec x}{dt}\Bigr)^2$.  The action can be rewritten so that it possesses a reparametrization symmetry.  Following \cite{Banerjee:2004ms}, one can preserve the Galilei invariance, while introducing a gauge degree of freedom $t(\lambda)$ associated with the time [where $\lambda$ is an evolution parameter with no a priori relation to $t$]. The corresponding action is
\be S_{\rm nr}=  \int d\lambda\;\frac m2 \frac {\dot{\vec x}^2}{\dot t} \;, \label{rpnvrntnrctn}\ee
the dot denoting differentiation in $\lambda$.
If $p_i$ denotes the momenta conjugate to $x_i$ and $\pi_t$ denotes the momentum conjugate to $t$ one gets 
\be p_i=m\frac{\dot x_i}{\dot t}\qquad\qquad\pi_t=-\frac m2\,\frac{\dot{\vec  x}^2}{\dot t^2}\;,\ee and hence 
the Hamiltonian constraint 
\be \Psi_1=\vec p^2 +2m\pi_t \approx 0 \;,\ee
generating the reparametrization symmetry $\lambda\rightarrow \lambda'(\lambda)$.
The symmetry is eliminated upon identifying $\lambda$ with the time $t$, in which case (\ref{rpnvrntnrctn}) reduces to  $S^0_{\rm nr}$.   Here let us instead use (\ref{ncnreltime}) to relate $\lambda$ to the time $t$.  That is, we impose the gauge constraint \be \Psi_2=t-\lambda -\frac {m\vec x\cdot \vec p}{\vec{ p}^2+\Lambda^2} \approx 0\label{nrfprtgfx} \ee
It is straightforward to show that the resulting Dirac brackets for $x_i$ and $p_i$ are (\ref{classsnydr}), yielding  the Euclidean Snyder algebra. The Dirac brackets of the time $t$ with $x_i$ and $p_i$ are
\be  \{x_i,t\}_{DB} =\frac{mx_j}{\Lambda^2}\Bigl(\delta_{ij} -\frac {2p_ip_j}{\Lambda^2+\vec p^2}\Bigr)\qquad\qquad\{p_i,t\}_{DB} =-\frac{mp_i}{\Lambda^2}\label{dbnrxtxp}
 \ee
Since the action (\ref{rpnvrntnrctn}) is Galilei invariant and leads to the Snyder algebra when imposing the gauge fixing (\ref{nrfprtgfx}), Galilei invariance must follow when defining the dynamical time for the Snyder algebra according to (\ref{ncnreltime}).

Next we promote the classical time  $t$ to a Hermitean operator  $\hat t_{\rm nr}$.
 Up to operator ordering ambiguities, it   is
\be  \hat t_{\rm nr}=\lambda+\frac 12 \Bigl(\hat x_i \hat p_i f_{\rm nr}(\vec{\hat  p}^2)+  f_{\rm nr}(\vec{\hat  p}^2)\hat p_i \hat x_i \Bigr)\;,\label{nrtmoprtr}\ee  
where
\be 
 f_{\rm nr}(\vec{\hat  p}^2)=m(\Lambda^2 +\vec{  p}^2)^{-1}\label{nrfrmlfrfnt} \ee  
Since the time operator depends on $\hat x_i$ and $\hat p_i$, it is dynamically determined. From (\ref{snydrsubalg}), we  derive the following commutation properties for the dynamical time operator: 
\beqa [ \hat x_i,\hat t_{\rm nr}]&=&\frac i{\Lambda^2}\Bigl(m \hat x_i  -{\hat x_j \hat p_j\hat p_i} f_{\rm nr}(\vec{\hat  p}^2) - f_{\rm nr}(\vec{\hat  p}^2){\hat p_i\hat p_j \hat x_j }\Bigr)\cr & &\cr[ \hat p_i,\hat t_{\rm nr}]&=&-\frac {im }{\Lambda^2}\,\hat p_i\label{nrcrfortm}\;,\eeqa which are the quantum analogues of (\ref{dbnrxtxp}).
It follows that $\hat t_{\rm nr}$ is rotationally invariant, $[ \hat L_i,\hat t_{\rm nr}]=0$.   The commutation relations (\ref{nrcrfortm}) differ from those postulated by Snyder for $\hat x^0$  in  (\ref{lrntzcvrnt}), which is obvious since Lorentz covariance is not present in (\ref{nrcrfortm}). 

The quantum analogue of (\ref{vecG}) is
\be \hat G_i=\frac 12(\hat t_{\rm nr}\hat p_i+\hat p_i\hat t_{\rm nr}) -m\hat x_i  \ee  $\hat G_i$ satisfy the algebra of Galilei boosts
\beqa  [\hat G_i,\hat p_j]&=&-im\delta_{ij}\cr &&\cr
[\hat G_i,\hat L_j]&=&i\epsilon_{ijk}G_k \cr
&&\cr [\hat G_i,\hat H_0]&=&-i\hat p_i\label{lgbrgbs}
\;\eeqa
Moreover, the full  Galilei algebra is recovered upon including the generators $\hat p_i,\hat L_i$ and 
the Hamiltonian
\be  \hat H_0=\frac{\hat p_i\hat p_i}{2m}\; \label{nnrltvstchmct}\ee
Here, in addition to (\ref{lgbrgbs}), one has the  commutation relations
\beqa [\hat L_i,\hat p_j]&=&i\epsilon_{ijk} p_k \cr &&\cr
 [\hat L_i,\hat L_j]&=&i\epsilon_{ijk}\hat L_k \;,\label{Lpcrs}\eeqa
and $[\hat p_i, \hat p_j]=[\hat p_i, \hat H_0]=[\hat L_i, \hat H_0]=
[\hat G_i,\hat G_j]= 0$.   
  The Galilei group  acts unitarily in the quantum theory.  The unitary actions of $\hat p_i$ and $\hat L_i$ were given explicitly in \cite{us}, and using them one can construct the remaining Galilei transformations.  This follows because the remaining generators $\hat G_i$ and $\hat H_0$ are written in terms of $\hat x_i$ and $\hat p_i$ which have a well-defined action on the states.

As  $\hat t_{\rm nr}$ is noncommuting, care must be taken in writing down a time-evolution operator in the quantum system. It is  constructed using the Hamiltonian  (\ref{nnrltvstchmct}) according to
 \be\tilde U(\lambda,0)= e^{-i \hat H_0\lambda}\label{evlutnnct}\ee  The evolution of the mean values of the space and time coordinates is determined from
\beqa \frac d{d\lambda} <\hat x_i>&=& -i<[\hat x_i,\hat H_0]> \;=\;
\Bigl<\frac{\hat p_i}{m}\Bigl(1+\frac{\vec{\hat p}^2}{\Lambda^2}\Bigr) \Bigr> \cr & &\cr  \frac d{d\lambda} <\hat t_{\rm nr}>&=& -i<[\hat t_{\rm nr},\hat H_0 ]> \; +\;\Bigl<\frac {\partial \hat t_{\rm nr}}{\partial\lambda}\Bigl> \;\;=\;\Bigl<1+\frac{\vec{\hat p}^2}{\Lambda^2}\Bigr >\label{lmbdevlxt}\;,\eeqa
and so the mean velocity is 
\be \frac{ d<\hat x_i>}{d<\hat t_{\rm nr}>} = \frac{\Bigl<\frac{\hat p_i}{m}\Bigl(1+\frac{\vec{\hat p}^2}{\Lambda^2}\Bigr) \Bigr> }{\Bigl<1+\frac{\vec{\hat p}^2}{\Lambda^2}\Bigr> } \label{meanvelnr}  \;,\ee
in contrast to the usual result (\ref{stndrdxpcxnr}).
The expression (\ref{meanvelnr}) reveals  the unusual feature that the mean velocity  is not linear in the  momentum, and  that the size and shape of the wavepacket can effect its mean velocity when one approaches the noncommutative scale $\Lambda$.  To illustrate this point we can consider a Gaussian distribution in momentum space  centered about $\vec p=(0,0,<p>)$,
\be \psi(\vec p)|_{\lambda=0}= \frac 1 {\pi^{3/4} \sigma^{3/2}}\; \Bigl(1+ \frac{ \vec p^2}{\Lambda^2}\Bigr)\exp{\biggl\{-\frac{p_1^2+p_2^2 +(p_3\;-<p>)^2}{2\sigma^2}}\biggr\}\;\label{gausswvpkt}
\ee 
The factor $1+ \frac{ \vec p^2}{\Lambda^2}$ was included to cancel out the momentum-dependent factor in the measure (\ref{measure}), and we assume $\sigma>0$.
The width  in momentum space $\Delta  p_i$ is  $\sigma /{\sqrt {2}}$ in any direction, $i=1,2,3$.  Using the differential representation for $\hat x_i$ in (\ref{noncandifrep}), we get that the  wavepacket is centered about the origin in position space when $\lambda=0$, $<\hat x_i>|_{\lambda=0}=0$.  For the width of the wavepacket in the $x_3$ direction, we find
\be (\Delta x_3|_{\lambda=0})^2  =\frac 1{{2}\sigma^2}\biggl\{ \Bigl(1+\frac{<p>^2}{\Lambda^2}\Bigr)^2+\frac{\sigma^2}{\Lambda^2}\Bigl(1+7\frac{<p>^2}{\Lambda^2}\Bigr)+\frac{11\sigma^4}{4\Lambda^4}\biggr\}\;\label{stcdltxthr}\;,
\ee    which reduces to the usual value  of $ 1/{\sqrt{2}\sigma}$ in the $\Lambda\rightarrow\infty$ limit. 
The effect of noncommutativity is to increase the spread of the wavepacket in position space. This  result  is consistent with the modified Heisenberg uncertainty relation.  From (\ref{stcdltxthr}) one can check that
\be   \Delta x_3  \Delta p_3\;\ge \; \frac 12 \Bigl(1+ \frac{ < p_3^2>}{\Lambda^2}\Bigr)\;, \ee which follows from the commutation relations (\ref{snydrsubalg}), is satisfied at $\lambda=0$.\footnote{We have assumed $\hbar=1$ throughout this article. For more discussion of  such uncertainty relations, see \cite{Kempf:1994su},\cite{Battisti:2008xy}. Here we get an additional  time-energy uncertainty relation  
$$ \Delta t_{\rm nr}\; \Delta\hat H_0 \;\ge\;\frac m{\Lambda^2}\; <\hat H_0>\;,$$ which follows from  
$ [\hat t_{\rm nr},\hat H_0] =\frac i{\Lambda^2}\;\vec {\hat p}^2 $.  For the wavepacket (\ref{gausswvpkt}), one gets 
$<\hat t_{\rm nr}>|_{\lambda=0}=0$ and 
 $$ (\Delta {\hat t_{\rm nr}}|_{\lambda=0})^2=\frac{m^2}{2 \Lambda ^4 } \left(3+\frac{<p>^2}{\sigma ^2} \right)\;,$$ after applying the differential representation (\ref{noncandifrep}).
  This along with 
$$  <\hat H_0>=\frac 1 {2m}\Bigl(<p>^2+\frac{3 \sigma ^2}{2}\Bigr) \;,\qquad  \Delta\hat H_0=\frac \sigma {2m}\sqrt{\frac{3 \sigma ^2}{2}+2 <p>^2 }$$ is consistent with the inequality.} 
At an arbitrary $\lambda$, we get from (\ref{lmbdevlxt}) that 
\beqa <\hat x_3>&=& 
\Bigl<\frac{\hat p_3}{m}\Bigl(1+\frac{\vec{\hat p}^2}{\Lambda^2}\Bigr) \Bigr>  \lambda \;=\; \Bigl(1+\frac{<p>^2 +\frac 52 \sigma^2}{\Lambda^2}\Bigr)\;\frac {<p>}m\;\lambda \cr & &\cr
<\hat t_{\rm nr}>&=&\Bigl<1+\frac{\vec{\hat p}^2}{\Lambda^2} \Bigr>\lambda\;=\;\Bigl(1+\frac{<p>^2 +\frac 32 \sigma^2}{\Lambda^2}\Bigr)\lambda\; \eeqa
 The resulting   mean speed of the Gaussian wavepacket as a function of its width and of $<p>$ is   
\be \frac{ d<\hat x_3>}{d<\hat t_{\rm nr}>} \;= \;\frac{2 +\frac{ 5 \sigma^2}{<p>^2 +\Lambda^2}} {2 +\frac{ 3 \sigma^2}{<p>^2 +\Lambda^2}}\;\;\frac {<p>}m\;\ee
For $\sigma<<\Lambda$ or  $|<p>|$, the  mean speed differs from ${<p>}/m$
 by a the factor of $1+{  \sigma^2}/{(<p>^2 +\Lambda^2)}$.

Unlike the position operators, the  time operator (\ref{nrtmoprtr}) has continuous eigenvalues.  We denote the latter by $\tau$. Then using the differential representation (\ref{noncandifrep}), we can solve the eigenvalue equation
\be \hat t_{\rm nr}\;\phi_\tau(\vec p,\lambda)= \tau\;\phi_\tau(\vec p,\lambda)\;\label{timeeigenvlu}\ee for the eigenfunctions in momentum space $\phi_\tau(\vec p,\lambda)$,
evaluated at some  $\lambda$.
 We get
\be  \phi_\tau(\vec p,\lambda)=\frac {\Lambda^2+p^2}{8\pi^2\Lambda^{2} m} \;\;p^{\frac{i\Lambda^2
(\lambda-\tau)}m-\frac 32}\label{nctmeigenfn}
  \ee    
 Their normalization is such that 
\be \int d\mu(\vec p) \;\phi_{\tau'}(\vec p,\lambda)^* \phi_\tau(\vec p,\lambda)=\delta(\tau-\tau')\label{orthotimfn}
\ee  To see this,  substitute (\ref{nctmeigenfn}) in the left-hand side  and apply the integration measure (\ref{measure}) to get
\be \frac {\Lambda^{2}}{2\pi m} \int^\infty_0  {dp} \;p^{\frac{i\Lambda^2
(\tau'-\tau)}m-1}\;
\ee
Then after making the change in integration variables from $p$ to $w=\frac {\Lambda^2}m \ln p  $,  this becomes $ \frac 1{2\pi}  \int^\infty_{-\infty}  {dw} \;e^{i
(\tau'-\tau)w}= \delta(\tau-\tau')$.

\subsection{Relativistic free particle}

We now repeat the previous analysis for the relativistic case.  $\lambda$ is  again given by   (\ref{lmdasntgrl}) along any trajectory, where  
 $C$ is some function of the constants of motion.   From   the  equation of motion $ {d\vec x}/{dt}=\vec p/p^0$, the constants of motion are now $\vec p$, $\vec L$ and \be \vec K=t\vec p-\vec x p^0\;\label{vecK}\ee We again define $p^0$ using the mass shell condition (\ref{pzerocv}).
The full Poincar\'e symmetry of the relativistic particle is recovered for the choice $C=\vec p \cdot\vec K$, and hence
  \be  t=\lambda+\frac{p^0\;\vec x\cdot \vec p}{\Lambda^2 +\vec p^2}  \label{gfxng}\;,\ee with the constants of motion generating the symmetry.  
 Eq. (\ref{gfxng}), along with the equations of motion, leads to
\be \frac{dx^i}{d\lambda}=\frac{p_i}{p^0}\Bigl (1+\frac{\vec p^2}{\Lambda^2}\Bigr)\qquad\qquad \frac{dt}{d\lambda}=1+\frac{\vec p^2}{\Lambda^2}\;\label{drvstlmbd}\ee The second equation indicates that $\lambda$ is a momentum-dependent rescaling of the time $t$ on any particle world line, as noted previously. Eq.
(\ref{gfxng}) reduces to (\ref{ncnreltime}) in the nonrelativistic limit. 

  Analogous to what happens in the nonrelativistic case, Poincar\'e symmetry  follows because one can write down a reparametrization invariant particle action, which yields the Snyder algebra when imposing (\ref{gfxng}) as a gauge fixing.\cite{Stern:2010ri}
The action can be taken to be the standard one for a
 relativistic free particle action
 \be
S_{\rm rel} =-m\int d\lambda \;\sqrt{-\dot  x_\mu\dot x^\mu}\label{rltscactn}\;,\ee  with the dot again denoting  differentiation in evolution parameter $\lambda$  [which we assume has no a priori relation to $x^0$].  As is well known, the action leads to the mass shell constraint \be
\Psi_1=p^\mu p_\mu +m^2\approx 0 \;,\ee in the Hamiltonian formalism, with $p_\mu$ canonically conjugate to $x^\mu$.  The constraint generates the reparametrization symmetry, which may be eliminated by imposing an additional constraint on the time $t=x^0$.  It was shown in \cite{Stern:2010ri}, that the Euclidean Snyder algebra (\ref{classsnydr}), is realized by the Dirac brackets if one instead chooses the gauge fixing condition to be
\be \Psi_2=  t-\lambda-\frac{p^0\;\vec x\cdot \vec p}{\Lambda^2 +\vec p^2}\approx 0 \label{gaugecond2}\;\ee
Since the action (\ref{rltscactn}) is Poincar\'e invariant and leads to the Snyder algebra when imposing this gauge fixing (\ref{gaugecond2}), Poincar\'e invariance must  follow when defining the dynamical time for the Snyder algebra according to (\ref{gfxng}).  Below we shall show explicitly that this symmetry is generated by  $ p_\mu$, $ L_i$ and $K_i$.

Up to operator ordering ambiguities, the  Hermitean operator  $\hat t_{\rm rel}$  associated with $t$ in (\ref{gfxng}) is
\be  \hat t_{\rm rel}=\lambda+\frac 12 \Bigl(\hat x_i \hat p_i f_{\rm rel}(\vec{\hat  p}^2)+  f_{\rm rel}(\vec{\hat  p}^2)\hat p_i \hat x_i \Bigr)\;,\label{nrtmoprtrel}\ee   where  the function   $f_{\rm rel}((\vec{\hat  p}^2)$ is  given by
\be 
 f_{\rm rel}(\vec{\hat  p}^2)=(\Lambda^2 +\vec{\hat  p}^2)^{-1}\hat p^0 \label{tmoprtr}\;,\ee  and  reduces to (\ref{nrfrmlfrfnt}) in the nonrelativistic limit.
  $\hat t_{\rm rel}$ is dynamically determined since it is a function of $\hat x_i$ and $\hat p_i$.  It is inequivalent to the time operator $\hat x^0$ appearing in
(\ref{lrntzcvrnt}).   Unlike (\ref{lrntzcvrnt}), the algebra generated by  $\hat t_{\rm rel}, \hat x_i, \hat p_i$ and $\hat p^0$ [given in (\ref{pzero}))]
  is not Lorentz covariant.  From (\ref{snydrsubalg}), we get the  commutation relations 
\beqa [ \hat x_i,\hat t_{\rm rel}]&=& \frac i{2\Lambda^2}\biggl(\hat p^0  \hat x_i\; +\;  \hat x_i \hat p^0\; +\;\frac{\Lambda^2-(\hat p^0)^2-m^2}{\Lambda^2 +\vec{\hat  p}^2}\;\frac {\hat p_i \hat p_j}{\hat p^0 }\hat x_j\;+\;\hat x_j \frac{\hat p_j\hat p_i}{\hat p^0 }\;\frac{\Lambda^2-(\hat p^0)^2-m^2}{\Lambda^2 +\vec{\hat  p}^2} \biggr)\cr & &\cr[ \hat p_i,\hat t_{\rm rel}]&=&-\frac {i }{\Lambda^2}\,\hat p^0\hat p_i\cr & &\cr[ \hat p^0,\hat t_{\rm rel}]&=&-\frac {i }{\Lambda^2}\,\hat p_i\hat p_i\;\label{tctsrel}\eeqa
The second equation agrees with  Snyder's commutator $[\hat p_i,\hat x^0]$, while the remaining two equations differ.  The third equation follows from the second using (\ref{pzero}).  The time operator is once again rotationally invariant,  $[ \hat L_i,\hat t_{\rm rel}]=0$. 

We recover the Poincar\'e algebra, despite the lack of covariance.  The quantum analogue of (\ref{vecK}) is
\be \hat K_i=\frac 12(\hat t_{\rm rel}\hat p_i+\hat p_i\hat t_{\rm rel}-\hat x_i \hat p^0-\hat p^0\hat x_i)\;\ee From it we obtain the standard  commutation relations for the Lorentz boosts,
\beqa [\hat K_i, \hat p^0]&=& -i\hat p_i \cr &&\cr
 [\hat K_i, \hat p_j]&=& -i\delta_{ij}\hat p^0
\cr &&\cr
 [\hat K_i, \hat L_j]&=& i\epsilon_{ijk}\hat K_k\cr &&\cr
 [\hat K_i, \hat K_j]&=&- i\epsilon_{ijk}\hat L_k\;
 \eeqa
Upon including $\hat p_i, \hat p^0$ and $\hat L_i$, we obtain
(\ref{Lpcrs})
and $[\hat p_i, \hat p_j]=[\hat p_i, \hat p^0]=[\hat L_i, \hat p^0]=0$, and hence the Poincar\'e algebra.
The Hilbert space for the theory therefore carries unitary representations of Poincar\'e group.

 Poincar\'e transformations of space-time are implemented  with the action of $\hat p_i$, $\hat p_0$, $\hat L_i$ and $\hat K_i$.  To see this we need to insure that the gauge condition (\ref{gfxng}) is preserved while also performing a reparametrization $\lambda\rightarrow \lambda'(\lambda)$ in the action (\ref{rltscactn}).   One implements the Poincar\'e transformations on space-time, as usual, using commutators with the generators, only here the position operators have discrete spectra.  Below we simplify to the classical theory and show that infinitesimal space translations, time translations, rotations and Lorentz boosts are obtained by taking Poisson brackets, respectively, with  the Poincar\'e generators $ p_i$, $ p_0$, $ L_i$ and $ K_i$. 
\begin{enumerate}
\item Infinitesimal spatial translations leave the four-momenta invariant, while the space-time coordinates of the particle $x^i(\lambda)$ and $t(\lambda)$ undergo variations
\be \delta x^i(\lambda) = \epsilon^i + \frac{dx^i}{d\lambda}\delta\lambda\qquad\qquad  \delta t(\lambda) =  \frac{dt}{d\lambda}\delta\lambda\;, \ee
$\epsilon_i$ being infinitesimal parameters.  It remains to find how $\delta\lambda$ depends on $\epsilon^i$. Using (\ref{drvstlmbd}),
\be \delta x^i(\lambda) = \epsilon^i +\frac{p_i}{p^0}\Bigl (1+\frac{\vec p^2}{\Lambda^2}\Bigr)\delta\lambda\qquad\qquad  \delta t(\lambda) = \Bigl (1+\frac{\vec p^2}{\Lambda^2}\Bigr) \delta\lambda \label{nflsptltrn}\ee The infinitesimal reparametrization  $\delta\lambda$ is constrained by the requirement that  the gauge condition  (\ref{gfxng}) is preserved by the variations (\ref{nflsptltrn}), i.e.,
 \be  t+\delta t(\lambda)=\lambda+\frac{p^0\;\Bigl(\vec x+\delta\vec x(\lambda)\Bigr)\cdot \vec p}{\Lambda^2 +\vec p^2} \ee   This gives
\be \delta\lambda=\frac{p^0\;\vec\epsilon\cdot \vec p}{\Lambda^2 +\vec p^2}\ee
It then follows that the infinitesimal variations  (\ref{nflsptltrn}) associated with spatial translations are obtained from Poisson brackets with $p_i$,
\beqa \delta x^i(\lambda)& = &    \Bigl(\delta_{ij} +\frac{p_ip_j}{\Lambda^2}\Bigr)\epsilon_j\;=\;\{x^i,\vec \epsilon\cdot\vec p\}\cr &&\cr  \delta t(\lambda) &=&\frac{p^0p_i}{\Lambda^2}\epsilon_i\;=\;\{t,\vec \epsilon\cdot\vec p\}
\eeqa
\item Infinitesimal time translations also  leave the four-momenta invariant, while
\be \delta x^i(\lambda) =\frac{p_i}{p^0}\Bigl (1+\frac{\vec p^2}{\Lambda^2}\Bigr)\delta\lambda\qquad\qquad  \delta t(\lambda) =\epsilon^0+ \Bigl (1+\frac{\vec p^2}{\Lambda^2}\Bigr) \delta\lambda\;, \label{nfltimetrn}\ee 
where $\epsilon^0$ is infinitesimal and we again used   (\ref{drvstlmbd}).   In this case the  gauge condition  (\ref{gfxng}) is preserved for 
\be \delta\lambda= -\epsilon^0\;,\ee and  the infinitesimal variations  (\ref{nfltimetrn}) associated with time translations are obtained from Poisson brackets with $p_0$,
\beqa \delta x^i(\lambda)& = &   - \frac{p_i}{p^0}\Bigl (1+\frac{\vec p^2}{\Lambda^2}\Bigr)\epsilon^0\;=\;\{x^i, \epsilon^0 p_0\}\cr &&\cr  \delta t(\lambda) &=&-\frac{\vec p^2}{\Lambda^2}\epsilon^0\;=\;\{t,\epsilon^0p_0\}
\eeqa
\item Infinitesimal rotations leave $p_0$ and $t$ invariant, while
\be\delta x^i=\epsilon_{ijk} \eta^jx^k\qquad\qquad \delta p_i=\epsilon_{ijk}\eta^j p_k\;,\label{nftlmrtns}\ee  where $\eta^k$ are infinitesimal. The transformation involves no reparametrizations since  the  gauge condition  (\ref{gfxng}) is rotationally invariant. As usual, (\ref{nftlmrtns}) can be expressed in terms of  Poisson brackets with $L_i$,
\be\delta x^i=\{x^i,\vec \eta\cdot\vec L\}\qquad\qquad \delta p_i=\{p_i,\vec \eta\cdot \vec L\}\ee
\item
Finally, infinitesimal Lorentz boosts are of the form
\beqa \delta x^i(\lambda) =\omega^{0i}t(\lambda) +\frac{p_i}{p^0}\Bigl (1+\frac{\vec p^2}{\Lambda^2}\Bigr)\delta\lambda\quad &\;&\quad  \delta t(\lambda) = \omega^{0i}x^i(\lambda) +\Bigl (1+\frac{\vec p^2}{\Lambda^2}\Bigr) \delta\lambda\cr &&\cr  \delta p_i =- \omega^{0i}p_0\qquad\quad\quad\qquad\qquad\qquad &\;&\quad\delta p_0 =- \omega^{0i}p_i\;,\label{nfntllrntzbst}\eeqa 
where $\omega^{i0}$ are infinitesimal.  In order for  the  gauge condition  (\ref{gfxng}) to be preserved, one now needs
\be \delta \lambda=\frac{-\omega^{0i}}{\Lambda^2+\vec p^2}\biggl\{(\Lambda^2-m^2)x^i+\Big((t-\lambda)(m^2-\Lambda^2)-\lambda(p^0)^2\Bigr)\frac{p_i}{p^0}\biggr\}
\ee  The infinitesimal Lorentz boosts (\ref{nfntllrntzbst}) can then be obtained by taking Poisson brackets  with $K_i$ defined in (\ref{vecK}),
\beqa \delta x^i(\lambda) &=&\omega^{0j}\biggl\{t\delta_{ij}-\frac{p_ix_j}{p^0}+\frac {p_ip_j}{\Lambda^2}\Bigl(\lambda +\frac{(\Lambda^2 -m^2)\,(t-\lambda)}{(p^0)^2}\Bigr)+\frac{m^2}{\Lambda^2}\frac{p_i x_j}{p^0}\biggr\}= \{x_i,\omega^{0j}K_j\}\cr&&\cr
\delta t(\lambda)& =& \frac{\omega^{0i}}{\Lambda^2}\biggl\{ {m^2}\,x_i + {p_ip^0}\Bigl( \lambda +
\frac{(\Lambda^2-m^2)(t-\lambda)}{(p^0)^2}\Bigr)\biggr\}=\{t,\omega^{0i}K_i\}\;,
\eeqa 
in addition to $\delta p_\mu =\{p_\mu,\omega^{0i}K_i\}$.
\end{enumerate}

  The evolution operator (\ref{evlutnnct}) can be applied to the relativistic system upon choosing $\hat H_0$ equal to  $\hat p^0$.  Then 
 the mean values of $\hat x_i$ and $\hat t$ evolve according to
\beqa \frac d{d\lambda} <\hat x_i>&=& -i<[\hat x_i,\hat p^0]> \;=\;
<\frac{\hat p_i}{\hat p^0}\Bigl(1+\frac{\vec{\hat p}^2}{\Lambda^2}\Bigr) >   \cr & &\cr  \frac d{d\lambda} <\hat t_{\rm rel}>&=& -i<[\hat t_{\rm rel},\hat p^0]> \; +\;\Bigl<\frac {\partial \hat t_{\rm rel}}{\partial\lambda}\Bigl>\;\;\;=\;<\Bigl(1+\frac{\vec{\hat p}^2}{\Lambda^2}\Bigr) >\;,\eeqa 
giving a mean velocity of
 \be \frac{ d<\hat x_i>}{d<\hat t_{\rm rel}>} = \frac{<\frac{\hat p_i}{\hat p^0}\Bigl(1+\frac{\vec{\hat p}^2}{\Lambda^2}\Bigr) > }{<\Bigl(1+\frac{\vec{\hat p}^2}{\Lambda^2}\Bigr)> }  \;\ee  This is in contrast with the result in (\ref{dmnxdt}).  So, as in the nonrelativistic case, the mean speed depends on the size and shape of the wavepacket as one approaches the noncommutative scale.
  
 Also as in the nonrelativistic case, the dynamical time operator (\ref{nrtmoprtrel})  has continuous eigenvalues which we again denote by $\tau$.  Using the differential representation (\ref{noncandifrep}),    the momentum-dependent eigenfunctions of $\hat t_{\rm rel}$ are
\be  \phi_\tau(\vec p,\lambda)={\cal M}e^{i\kappa\tau}\;\frac{\Lambda^2+p^2}{p^{3/2} (p^0)^{1/2}}\;\biggl\{\frac m{p\Lambda}(m+p^0) \biggr\}^{\frac{i\Lambda^2
(\tau-\lambda)}m}\label{tmeigenfn}
  \ee
 The  factor ${\cal M}e^{i\kappa\tau}$ is determined from the orthonormality condition (\ref{orthotimfn}).
After substituting in (\ref{tmeigenfn}) and applying the integration measure (\ref{measure}), the left-hand side of  (\ref{orthotimfn}) becomes
\be 4\pi \Lambda^{4} |{\cal M}|^2 e^{i\kappa(\tau-\tau')}\int^\infty_0 \frac {dp}{pp^0} \;\biggl\{\frac m{p\Lambda}(m+p^0) \biggr\}^{\frac{i\Lambda^2
(\tau-\tau')}m}\;,
\ee
which after a change of integration variables from $p$ to $w=\kappa+\frac {\Lambda^2}m \ln \Bigl\{\frac m{p\Lambda}(m+p^0) \Bigr\} $,  simplifies to
\be 4\pi \Lambda^{2} |{\cal M}|^2 \int^\infty_{\kappa+\frac {\Lambda^2}m \ln \{\frac m{\Lambda} \} }  {dw} \;e^{i
(\tau-\tau')w}
\ee The delta function on the right-hand side of (\ref{orthotimfn}) is recovered for $ \kappa\rightarrow -\infty $ and $|{\cal M}| =  1/({2\sqrt{2}\pi \Lambda}) $.

\subsection{$1$D conservative systems}

As an example of an interacting system we examine the simplest case of a one-dimensional conservative system. Following the procedure of the previous examples, we  require the Hamiltonian generating dynamics on Snyder space to be identical to the Hamiltonian $H_0$ on canonical phase space,
\be H=H_0=\frac{p^2}{2m}+ V(x)\;,\ee
and so the evolution parameter $\lambda$ cannot be identified with the time $t$.  The relation between the two can once again be regarded as a gauge fixing condition.  For this we should start with an action which is reparametrization invariant.  Such an action can be written as
\be  S_{\rm con} = \int d\lambda \Bigl\{ \frac  m2 \,\frac{\dot x^2}{\dot t}-V(x)\dot t\Bigr\}\;,\ee
with the dot again denoting differentiation in the evolution parameter $\lambda$.  It reduces to the standard  action for a conservative system when $\lambda$ equals $t$.  The Hamiltonian constraint 
generating the reparametrization symmetry $\lambda\rightarrow \lambda'(\lambda)$ is now
\be \Psi_1= H_{\rm 0}+\pi_t \approx 0 \;,\ee  with $p$ and $\pi_t$  canonically conjugate to $x$ and $t$, respectively.  Since we are only considering one spatial dimension, the Snyder algebra in this case consists of the single  relation between $x$ and $p$,
\be \{x,p\}_{\rm DB}=1+\frac {p^2}{\Lambda^2}  \;,\ee
which we wish to have realized by the Dirac brackets.  Then if $\Psi_2\approx0$ denotes the gauge fixing condition its Poisson brackets should satisfy
\be \frac {p^2}{\Lambda^2} =\frac{\{x,\Psi_1\}\{\Psi_2,p\}-\{x,\Psi_2\}\{\Psi_1,p\}}{\{\Psi_1,\Psi_2\}} =\frac{-\{H_{\rm 0},\Psi_2\}}{\{H_{\rm 0}+\pi_t,\Psi_2\}}\;,\ee or simply
\be \{H_{\rm 0},\Psi_2\}=\frac{p^2}{p^2+\Lambda^2}\, \{\Psi_2,\pi_t\}\label{dfqfrpsi2}\ee

Explicit solutions for $\Psi_2$ depend on the form for the potential energy $V(x)$.  For the example of a  linear potential $V(x)=-Fx$, one has\be
\Psi_2=t+\frac 1F \Bigl(\Lambda\tan^{-1}\frac p\Lambda-p\Bigr)-\lambda\;,\label{gffrlnrptntl}\ee up to an additive function of $H_0$.  Setting $\Psi_2$ (strongly) equal to zero,  the evolution parameter $\lambda$ ranges over a finite domain of length $\pi\Lambda /F$ for any classical trajectory, $p=p_{\rm sol}= Ft +{\rm constant}$.  For such trajectories,
$\lambda$ and $t$ scale as in (\ref{heqlhzro}).  The  Dirac brackets of the time $t$ with $x$ and $p$ resulting from the gauge fixing (\ref{gffrlnrptntl}) are
\be 
\{t,x\}_{\rm DB}=- \frac {p^2}{F\Lambda^2}\qquad\qquad
\{t,p\}_{\rm DB}=0 \;,\ee
which then can be promoted to commutation relations in the quantum theory.

Another example is  the one-dimensional harmonic oscillator with the standard   Hamiltonian    (\ref{sho}).   Now (\ref{dfqfrpsi2}) 
 has the solution
\be \Psi_2=t\;+\;\frac 1{\omega}  \Bigl\{\alpha  \tan^{-1}{\alpha m\omega x/p} 
\; -\; \tan^{-1}{m\omega x}/p \Bigr\} \; -\;\lambda\;,\qquad\alpha=\frac 1{\sqrt{1+\frac{2m H_{\rm ho}}{\Lambda^2} }}\;,\ee up to an additive function of $H_0$.   This agrees with the result of the integration in  (\ref{ntgrlfrt}) along classical trajectories   $x=x_{\rm sol}=A\sin(\omega t+\phi)$, $\;p= p_{\rm sol}=m\omega A\cos(\omega t+\phi)$ .  
  Also, it reduces to the gauge fixing (\ref{nrfprtgfx}) for the  free nonrelativistic particle in the limit $\omega\rightarrow 0$. (This differs from the previous example where the free particle limit $F\rightarrow 0$ was singular.)
The resulting expressions for the Dirac brackets of the time $t$ with $x$ and $p$ are
\beqa  
\{t,x\}_{\rm DB}&=& \frac {p\alpha^2}{\omega\Lambda^4}\Bigl(m\omega x p - (\Lambda^2+p^2) \alpha\tan^{-1}{\alpha m\omega x/p}\Bigr)  \cr &&\cr 
\{t,p\}_{\rm DB}&=&\frac{m\alpha^2}{\Lambda^4} (\Lambda^2+p^2) \Bigl( p +  m\omega x  \alpha\tan^{-1}{\alpha m\omega x/p}\Bigr) \eeqa
It is a nontrivial problem to find a quantization of this system.  Also, the  generalization of  these examples to higher dimensions is not immediately obvious.

\section{Discussion}

In section 3, we  gave  conditions [Cf. (\ref{frstdrvHzro})] for writing down any given dynamical system on the symplectic manifold associated with  the Snyder algebra. Such a manifold is  characterized by  the symplectic two-form (\ref{smplct2frm}). Solutions, if they exist, may not be unique, and require fixing both the Hamiltonian $H$ and the evolution parameter $\lambda$ in terms of the phase space variables and the time.  Solutions were  found   for the case of free particle.  
There we could either set the time equal to the evolution parameter $\lambda$, so that it remains a real parameter in the quantum theory, or in an  alternative  approach, equal to a monotonically increasing function of $\lambda$  and the phase space variables.  In the latter, the time gets promoted to a Hermitean operator in the quantum theory.  The standard energy-momentum dispersion relations were preserved in the latter approach, and the full Poincar\'e (Galilei) symmetry group  for  the relativistic (nonrelativistic) particle was recovered. This was because the system could be derived from a  Poincar\'e (Galilei) invariant action.
The introduction of a time operator in quantum theory has nontrivial consequences, in particular with regards to causality violation.  Although causality violation is a well-known feature of relativistic quantum mechanics,  the situation here is more dramatic.  From the evolution operator (\ref{evlutnnct}), we can write down a Green function which takes a state with time eigenvalue $\tau'$, associated with evolution parameter  $\lambda'$, to a state with time eigenvalue $\tau$, associated with evolution parameter $\lambda$. It can be expressed as a function of $\tau$, $\tau'$ and $\lambda-\lambda'$ using the time eigenfunctions  $\phi_\tau(\vec p,\lambda)$ [eqs. (\ref{nctmeigenfn}) or (\ref{tmeigenfn})],
\be G(\tau,\tau',\lambda-\lambda')=\int   d\mu(\vec p) \;\phi_{\tau}(\vec p,\lambda)^*\phi_{\tau'}(\vec p,\lambda')\;e^{ -i(\lambda -\lambda') H_0 }\ee
  The Green function does vanish when $\tau-\tau'>0$ and $\lambda -\lambda'<0$, and so admits acausal time evolution in the quantum theory.  Causality violation in relativistic particle mechanics is standardly cured by writing down the corresponding quantum field theory. However, this is unlikely to be the case here since the resulting noncommutative field theory is nonlocal.

The question of whether or not there exist solutions to (\ref{frstdrvHzro}) when interactions are present can be nontrivial.  In Section 3.2, we saw that for the case of one-dimensional systems we cannot in general set the evolution parameter equal to the time.  So for  general interactions,  time is not associated with a commuting parameter in the quantum theory.  In Section 4.3, we examined the specific case of an interaction with an external potential in one dimension.  The action was written in a reparametrization invariant way and the time was  obtained from a gauge fixing condition.  Interactions with external potentials in higher dimensions should also be possible.  Of particular interest are interactions between relativistic particles, which have been examined in a general context in \cite{AmelinoCamelia:2011bm}.   A nontrivial co-product was introduced for the purpose of studying multiparticle states in a related work.\cite{Agostini:2003vg}
  On the other hand, the treatment of particle interactions (in the absence of  field theory) is problematic within our Hamiltonian framework,  due to no interaction  theorems.\cite{Currie:1963rw}
  Thus a reasonable direction for studying interactions for the  systems discussed here is to develop  field theory  on Snyder space. 
Preliminary attempts to write down field theory on Snyder space were given in \cite{Battisti:2010sr},\cite{Girelli:2010wi}, which rely on star product representations of the Snyder algebra.  However, the proposed star products have certain inconsistencies, such as non-associativity. Since Snyder space is, in fact, a quantum lattice, it makes sense to instead  consider a lattice field theory.  Our work with particle dynamics on the quantum lattice,  should serve as the ground work for writing fields on such a lattice.

\bigskip
{\Large {\bf Acknowledgments} }

\noindent
 This work was supported in part by the DOE,
Grant No. DE-FG02-10ER41714.

\end{document}